# OPERATIONS STRUCTURE FOR THE MANAGEMENT, CONTROL AND SUPPORT OF THE INFN-GRID/GRID.IT PRODUCTION INFRASTRUCTURE

Cristina Aiftimiei[a], Sabrina Argentati[b], Stefano Bagnasco[c], Alex Barchiesi[d], Riccardo Brunetti[c], Andrea Caltroni[a], Luciana Carota[e], Alessandro Cavalli[e], Daniele Cesini[e], Guido Cuscela[f], Marcio Da Cruz[c], Simone Dalla Fina[a], Cesare Delle Fratte[g], Giacinto Donvito[f], Sergio Fantinel[a], Federica Fanzago[a], Enrico Ferro[a], Sandro Fiore[i], Luciano Gaido[c], Francesco Gregoretti[l], Federico Nebiolo[c], Alfredo Pagano[e], Alessandro Paolini[e], Matteo Selmi[e], Rosario Turrisi[a], Luca Vaccarossa[h], Marco Verlato[a], Paolo Veronesi[e], Maria Cristina Vistoli[e]

*Istituto Nazionale di Fisica Nucleare (INFN), Sez. di Padova[a], Lab. Naz. di Frascati[b], Sez. di Torino[c], Sez. di Roma-1[d], CNAF-Bologna[e], Sez. di Bari[f], Sez. di Roma-2[g], Sez. di Milano[h], Italy*
*[i] University of Lecce and SPACI Consortium, Italy*
*[l] Istituto di Calcolo e Reti ad Alte Prestazioni-Consiglio Nazionale delle Ricerche (ICAR-CNR), Italy*

*Abstract*

Moving from a National Grid Testbed to a Production quality Grid service for the HEP applications requires effective operations structure and organization, proper user and operations support, flexible and efficient management and monitoring tools.

Moreover, easily deployable middleware releases should be provided to the sites, with flexible configuration tools suitable for heterogeneous local computing farms.

The organizational model, the available tools and the agreed procedures for operating the national/regional Grid infrastructures that are also part of the world-wide EGEE Grid play a key role for the success of a real production Grid, as well as the interconnection of the regional operations structures with the global management, control and support structure.

In this paper we describe the operations structure that we are currently using at the Italian Grid Operation and Support Centre.

The activities described cover monitoring, management and support for the INFN-GRID/GRID.IT production Grid (comprising more than 30 sites) and its interconnection with the EGEE/LCG structure as well as the roadmap to improve the global support quality, stability and reliability of the Grid service.

## THE NATIONAL GRID CENTRAL MANAGEMENT TEAM

*The Central Management Team*

The Central Management Team (CMT) is responsible for the registration procedure, middleware deployment and certification procedure for all INFN-GRID sites.

*Registration Procedure*: the Site Manager at the candidate site contacts the Italian ROC (Regional Operation Centre) CMT, giving contact information and service level definition numbers and a statement of acceptance of the policy documents. When the CMT agrees it is a good candidate site, the ROC operator creates the new site's record in the GOC (Grid Operation Centre [1]) database. At this stage the site status is marked as "candidate". The Resource Administrator at the candidate site enters the remaining information in the GOC database then requests validation by the ROC. If all required site information has been provided, the ROC manager changes the site status to "uncertified".

*Deployment Plan:* services deployment and Grid sites installation are scheduled according to a plan set forth by the Central Management Team. Accordance to the plan is mandatory to avoid having too many site upgrades during the day and mainly to ensure that the service level offered to users and Virtual Organisations (VOs) is acceptable even during the upgrade.

A deployment plan is published whenever a new middleware release is available.

Using the Italian ticketing system, site managers then publish downtime advices to the INFN-GRID calendar and, if a site is also registered in the EGEE GOC database, set downtime periods in the GOC database and announce it through the EGEE broadcast tool to all Virtual Organisation's members. The site's status is then changed to "closed" in the site Information System.

In a typical Grid site, installing a large number of computers with the same characteristics (e.g. a set of Worker Nodes) is a common issue. In this case the CMT provides the site an automatic method to setup the farm, where each node has the same basic configuration: the aim is to simplify the Grid middleware deployment in the INFN-GRID/GRID.IT infrastructure.

The procedure is provided through documents describing how to setup a server for RedHat OS installation via Kickstart using PXE technology and to manage a local APT repository for OS and middleware.

To manage the package repository a customized version of YAM, a tool to mirror RPM repositories [2], is used, while the OS installation is based on normal Kickstart files. The tool YAIM [3] is used to install/upgrade the middleware release.

*Site Certification Procedure:* having completed the procedures described above, the Resource Administrators at the site should:
- Apply for DTEAM and INFNGRID VO membership to allow test job submission to check the completeness of the local installation.
- Contact the CMT and ask for quality testing of the site installation.
- Request the CMT to perform acceptance tests before including the new site in the Information System. The site will be included in the Grid information system configuration files.

CMT then changes site status in the GOC database to "certified" and the site is included in the daily monitoring procedures.

## ROC shifts

The Italian ROC provides daily shifts to control and manage the Italian production Grid infrastructure. The shifts are organized with 2 people per shift and 2 shifts per day, Monday to Friday. The team on duty logs all the actions in a web-based logbook. The log is passed to the next shift to provide a summary of the current status of the Grid. The open/updated/closed trouble tickets, the status of Grid services and sites and the sites successful certifications are logged.
During the shift the certification procedure is started for those sites whose downtime advice expired.
GridICE, a grid monitoring tool developed in the EGEE/LCG framework [4], is used to check the status of the production Grid services and information systems in all the computing and storage resources.
Shifters on duty also use the Site Functional Tests page [5] to check the status of Italian production sites, and look at the GIIS status in all the production sites by accessing the LCG monitoring map [6] and the Italian top level BDII (Berkeley Database Information Index).

# USER REPORT AND ACCOUNTING

## The DGAS Accounting Service

The Distributed Grid Accounting Service (DGAS) [7], developed within European DataGRID project and EGEE project, implements a resource usage metering and economic accounting in a fully distributed Grid environment.

Each Grid computing resource and Grid user is registered in an appropriate server, known as HLR (Home Location Register), which keeps track of every submitted job. An arbitrary number of HLR servers can be used. By querying the HLR, it is possible to collect usage information on a single user or VO or group of users basis.

Usage records of each completed job are replicated both on "users" HLR and "resource" HLR. The first can be used by VO managers to collect information on user's activity, while the second can be queried by site managers to monitor usage of their computing resources.

The DGAS service has been included in the INFN-GRID middleware releases starting with version 2.6.0.

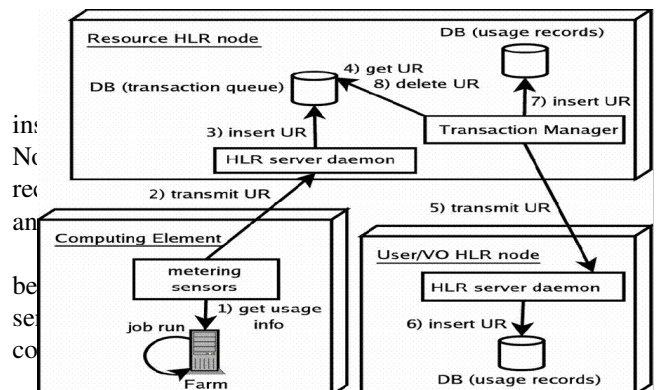

particular those responsible to collect usage records and to send information to the HLR; afterwards, a simple test job is submitted on each site of the production Grid. For sake of simplicity a single HLR is used for all the sites, but the target HLR can be easily specified inside the Job Description Language (JDL) file.

After the test is executed, the corresponding accounting information is checked, using the appropriate software client to query the users HLR.

This tool is currently used by ROC shifters during their daily activity to discover possible problems and to allow a quick fix.

## The reporting tool

In order to provide usage graphs, VO comparison metrics, global number of jobs executed in the production Grid infrastructure, a reporting tool called ROCRep has been developed [8].

It is possible to report the number of submitted jobs per site and per Virtual Organisation. ROCRep can generate either a tabular report with detailed information or graphical report based on pie charts and line graphs.

It is also possible to report the total Wall Clock Time and CPU Time used, again per site and per VO.

Furthermore, trend curves of executed jobs submitted in a selected set of sites are also available, making it possible, for example, to check whether during a specific day there have been problems or, in general, Grid resources have been overloaded or underexploited.

Another feature of ROCRep is a SLA (Service Level Agreement) service: thanks to this feature, it is possible to monitor whether the amount of resources provided by a site for the Grid remains stable. A chart is also provided for this purpose.

Reporting data currently originate from the GridICE monitoring tool database. In the future, starting with the INFN-GRID 2.7.0 deployment release, ROCRep will collect data directly both from the accounting system DGAS and from GridICE.

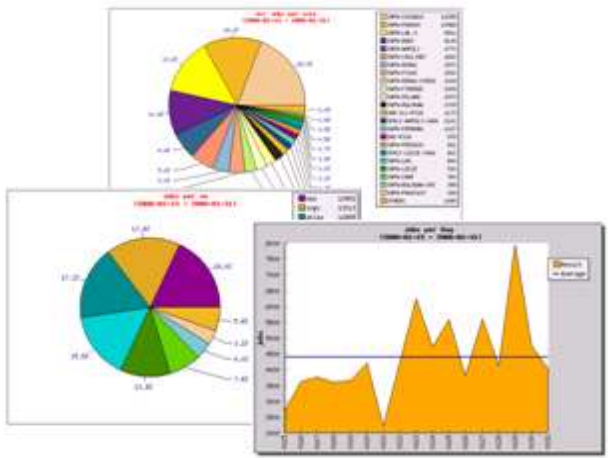

**Figure 2: ROCRep graphic reports**

## USER AND OPERATION SUPPORT

### Italian ROC helpdesk

Italian ROC ticketing system is built upon Xoops [9], a suite of web-based tools written in PHP. A complete portal with a large number of features is available for both Grid end-users and Resource Centres administrators. The former can consult the knowledge base or the Wiki to retrieve useful information and try to solve their problem by themselves; the latter can have fine-grained tuneable rights in order to give local support and to populate documentation sections.

XHelp is the main component of the suite, implementing a full helpdesk system with highly configurable ACLs (grouping users, site administrators, VO supporters, Central Management Team members, services technicians), unlimited custom fields (associating specific details with tickets), modifiable ticket workflow, per-user profile for tickets lists (searches can be saved and displayed in the summary page) and notification system (email, web-based messaging, RSS feed).

All components are accessible from the main interface of the portal, providing a single sign-on point of registration/identification certificate-based.

The end-user can open a request, view and follow his own tickets and related replies; a supporter can view tickets assigned to his own groups, add responses and solutions, and change status/priority; a ticket manager can moreover work with attachments, with FAQ and with ticket assignment (to other supporters and groups).

While operating on tickets, "side" content is always available for all classes of users (related to their access level): Site Functional Tests in RSS form, site downtimes calendaring system, file archive, net query tools, IRC applet, contextual questions and answers, reports from CMT daily shifts.

Administration of all the components is web-based too, enabling a restricted group of users to manage and modify behaviour of specific elements from a user-friendly GUI. The whole system is written in object oriented PHP, making it easy to extend single components, deactivate unneeded ones, hook modules among them, and add specific functionalities (like the GGUS interface described below). The whole suite is developed as an Open Source project, and specific ROC features are kept in sync with the official branch of the project, taking the advantages of both locally developed components and external updates and bug fixes.

### Integration with GGUS

Within EGEE, the ROC support infrastructures are connected via a central integration platform provided by the Global Grid User Support (GGUS) organisation. The GGUS model is described in detail in another paper presented at this conference [10].

Concerning the Italian ROC, the XHelp module described above has been interfaced to the GGUS helpdesk application using web-service technologies [11]. Secure methods to create and update trouble tickets in the GGUS database are provided by the GGUS application. These methods are called by APIs available in most programming languages, which wrap into SOAP messages the ticket information stored in the XHelp database, and send them to the WSDL contact URL. This way, a trouble ticket submitted by a local user to the XHelp helpdesk which could not be solved locally, can be escalated by the local supporter across the ROC boundaries. The system allows for ticket assignment to any other support unit of GGUS, as well as all other ROC helpdesks connected to GGUS via the interface. In this case, the ticket is shared among all the helpdesk's databases involved in the workflow, it can be updated from every source, and any updating action will propagate towards all the other systems.

On the other direction, trouble tickets originally submitted to GGUS can be assigned to the Italian ROC, triggering the automatic appearance of the ticket in the XHelp database, as well as the e-mail notification of the relevant support team. In this case, the GGUS ticket is sent to XHelp through an XML formatted e-mail, which is at first fetched from the mail server and then parsed by an importer tool running as a cron job on the XHelp server. The tool, written in Java, saves the ticket into the XHelp database, marking it as new if not existing yet locally, or updating an already existing ticket.

The possibility of replacing the importer tool with a web-services layer on top of XHelp, in order to allow synchronous transfer of tickets from GGUS to XHelp, in the same way as is done in the opposite direction, is under development and planned to be implemented within early months of 2006.

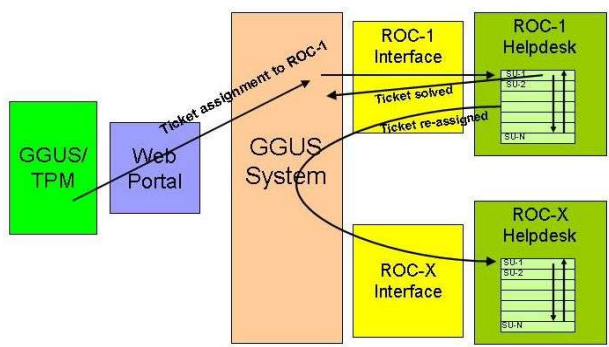

**Figure 3: Basic GGUS-ROC ticket workflow**

## ACKNOWLEDGEMENTS

We thank the Resource Centres administrators and the INFN-GRID/GRID.IT management:
Roberta Amaranti[1], Daniele Andreotti[1], Roberto Barbera[1], Alessandro Barilari[2], Massimo Biasotto[1], Tommaso Boccali[1], Giovanni Bracco[9], Alessandro Brunengo[1], Federico Calzolari[6], Antonio Cazzato[1], Andrea Chierici[1], Giovanni Ciraolo[1], Vitaliano Ciulli[1], Mirko Corosu[1], Marco Corvo[1], Stefano Dal Pra[1], Maurizio Davini[3], Luca Dell'Agnello[1], Alessandro De Salvo[1], Matteo Diarena[2], Flavia Donno[1], Alessandra Doria[1], Alessandro Enea[10], Italo Epicoco[8], Francesco Fabozzi[1], Alessandra Fanfani[1], Enrico Fasanelli[1], Maria Lorenza Ferrer[1], Vega Forneris[5], Antonio Forte[1], Riccardo Gargana[1], Osvaldo Gervasi[2], Riccardo Gervasoni[1], Antonia Ghiselli[1], Alessandro Italiano[1], Stefano Lusso[1], Marisa Luvisetto[1], Tullio Macorini[1], Giorgio Maggi[1], Nicolo' Magini[1], Mirco Mariotti[1], Fabio Martinelli[1], Enrico Mazzoni[1], Mirco Mazzucato[1], Massimo Mongardini[1], Daniele Mura[1], Igor Neri[1], Cristiano Palomba[1], Andrea Paolucci[1], Antonio Pierro[1], Lucio Pileggi[6], Giuseppe Platania[1], Silvia Resconi[1], Diego Romano[4], Davide Salomoni[1], Francesco Safai Tehrani[1], Franco Semeria[1], Marco Serra[1], Leonello Servoli[1], Antonio Silvestri[1], Alessandro Spanu[1], Lucio Strizzolo[1], Giuliano Taffoni[7], Francesco Taurino[1], Gennaro Tortone[1], Claudio Vuerli[7]

[1] Istituto Nazionale di Fisica Nucleare (INFN), Italy
[2] University of Perugia, Italy
[3] University of Pisa, Italy
[4] University of Naples, Italy
[5] ESA's European Space Research Institute (ESRIN), Italy
[6] Scuola Normale Superiore di Pisa, Italy
[7] Istituto Nazionale di Astrofisica (INAF), Italy
[8] University of Lecce and SPACI Consortium, Italy
[9] Ente per le Nuove tecnologie, l'Energia e l'Ambiente (ENEA), Italy
[10] Istituto di Linguistica Computazionale – Consiglio Nazionale delle Ricerche (ILC-CNR), Pisa, Italy

## REFERENCES

[1] http://goc.grid-support.ac.uk/gridsite/gocmain/
[2] http://dag.wieers.com/packages/yam/
[3] http://grid-deployment.web.cern.ch/grid-deployment/documentation/LCG2-Manual-Install/
[4] C. Aiftimiei, S. Andreozzi, G. Cuscela, N. De Bortoli, G. Donvito, S. Fantinel, E. Fattibene, G. Misurelli, A. Pierro, G.L. Rubini, G.Tortone, *"GridICE: Requirements, Architecture and Experience of a Monitoring Tool for Grid Systems",* to appear in Proceedings of the International Conference on Computing in High Energy and Nuclear Physics (CHEP2006), Mumbai, India, 13-17 February 2006
[5] https://lcg-sft.cern.ch:9443/sft/lastreport.cgi
[6] http://goc.grid.sinica.edu.tw/gstat/
[7] R. Piro, A. Guarise, and A. Werbrouck, *"An Economy-based Accounting Infrastructure for the DataGrid",* Proceedings of the 4th International Workshop on Grid Computing (Grid2003), Phoenix, Arizona, USA, November 17th, 2003

[8] http://grid-it.cnaf.infn.it/rocrep
[9] http://www.xoops.org/
[10] P. Strange, T. Antoni, F. Donno, H. Dres, G. Mathieu, A. Mills, D. Spence, M. Tsai and M. Verlato, *"Global Grid User Support: the model and experience in the Worldwide LHC Computing Grid"*, to appear in Proceedings of the International Conference on Computing in High Energy and Nuclear Physics (CHEP2006), Mumbai, India, 13-17 February 2006
[11] "GGUS-ROC Interface Project Home", http://infnforge.cnaf.infn.it/eticketimp